\documentclass{ws-ijmpb}
\usepackage{graphicx}% Include figure file
\newcommand{\figwidth}{5.0 in} % Use for single column output

\begin{document}

\markboth{K. Wierschem and E. Manousakis}{Monolayer charged quantum films}

\catchline{}{}{}{}{}

\title{Monolayer charged quantum films: A quantum simulation study}

\author{Keola Wierschem}
\address{Physics Department, Florida State University, Tallahassee, Florida, 32306-4350, keola@martech.fsu.edu}

\author{Efstratios Manousakis}
\address{Physics Department, Florida State University, Tallahassee, Florida, 32306-4350, stratos@martech.fsu.edu}

%%%%%%%%%%%%%%%%%%%%%%%%%%%%%%%%%%%%%%%%%%%%%%%%%%%%%%%%%%%%
% You may repeat \author \address as often as necessary    %
%%%%%%%%%%%%%%%%%%%%%%%%%%%%%%%%%%%%%%%%%%%%%%%%%%%%%%%%%%%%

\maketitle

\begin{history}
%\received{DAY MONTH YEAR}
%\revised{DAY MONTH YEAR}
\end{history}

\begin{abstract}
We use path-integral Monte Carlo (PIMC) to 
study the effects of adding a long-range
repulsive Coulomb interaction to the usual
Van der Waals interaction between two atoms of a submonolayer quantum
film such as helium on  graphite or a pure two-dimensional superfluid.
Such interactions  frustrate or compete  with the natural  tendency of
the system for phase separation namely to form a macroscopic liquid or
solid phase.  We  find that as a function of  the relative strength of
the long-range repulsion, surface coverage and temperature, the system
undergoes  a   series  of  transformations,   including  a  triangular
Wigner-like crystal of clusters, a charge stripe-ordered phase and a 
fluid phase.  The  goal of these
studies is  to understand the  role of quantum fluctuations  when such
competing interactions  appear together with  formation of preexisting
electron pairs as might be the case in cuprate superconductors.

\end{abstract}


\section{INTRODUCTION}
In a number of physical systems where there are competing interactions
such as a short-range attractive part and a long-range repulsion, a rich 
phase diagram is expected.
In the particular case where the long-range repulsion is the 
usual Coulomb interaction, nuclear matter and electrons in a solid such
as the cuprate superconductors are examples of this case. In 
the latter case, the electrons in the material can find themselves
experiencing an attractive interaction which might also lead to pairing
or phase separation, but when they form clusters
the accumulated charge cannot exceed a threshold value due to the 
Coulomb repulsion.

In this work we focus our attention to the case of a
two dimensional quantum fluid of charged bosons with short-range
attractive interaction and a hard core repulsion. 
This model, for example, could be used to understand
certain aspects of the conduction electrons of the copper-oxide plane
of the cuprate superconductors. It has been found\cite{emery,hm,carlson}
that in cuprate superconductors, as modeled by the
so-called $t-J$ model, holes tend to phase separate in a hole rich and
an antiferromagnetically ordered region. A pair of holes in a 
quantum antiferromagnet is strongly bound\cite{bm}
and in fact the hole rich phase can be understood as a fluid 
of weakly attracted bosons--the bound pairs of 
holes\cite{manousakis03}. In this paper, we consider such a system of
bosons which interact via a short range attractive interaction, a hard core 
repulsion and 
a long-range Coulomb repulsion. A system of interacting bosons
is free from the infamous minus problem which hinders the quantum simulation
of fermions. Therefore we should be able to address the phase diagram 
of this model without serious difficulties.

In the present paper our goal is  to understand the influence of a weak Coulomb
repulsion on a quantum Van der Waals-Bose liquid monolayer film. 
Our model of the neutral Van der Waals-Bose liquid 
is a liquid helium film, because  we have gained understanding
of the phases of this system from experimental 
studies\cite{greywall93,reppy93,reppy96}
and from our
previous PIMC studies\cite{pierce}.  More importantly, using the PIMC method
most of the observed features of these films and the phase diagram 
of each layer as a function of surface 
coverage can be reproduced with detail.
In these PIMC studies \cite{pierce} of submonolayer, second, third and 
higher layers of helium on graphite well-known helium-helium 
potential and helium-graphite interaction were used without the 
introduction of any fitting parameter. In this paper, we begin from 
the Hamiltonian
that describes  a submonolayer helium film on graphite and we modify it by 
adding  a weak Coulomb repulsion 
term to the helium-helium interaction potential which depends on 
a coupling constant $\alpha$.  We will use PIMC method
to study such a film as a function of temperature, surface density
and the parameter $\alpha$.

\section{THE APPROACH}

\begin{figure}
\begin{center}
\includegraphics[width=3.5 in]{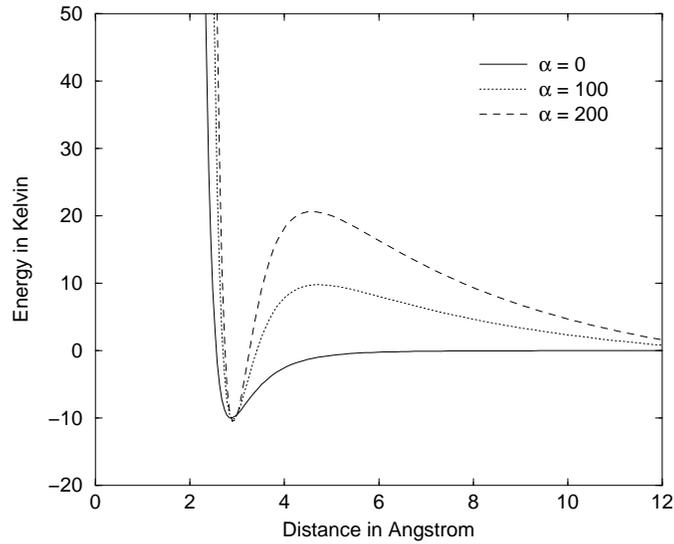}
\end{center}
\caption{The total interaction potential is shown 
for various values of the parameter $\alpha$. The value of $\epsilon$ is
adjusted such that the minimum of the
interaction is fixed at $-10^{\circ} K$.}
\label{fig0}
\end{figure}

For particle-substrate interaction we used the laterally averaged
helium-graphite interaction\cite{carlos} used in previous 
PIMC studies of helium adsorption on graphite\cite{pierce}.
In order to study the role of a long-range Coulomb repulsion on
the Van der Waals helium-helium atom interaction we added a
$1/r$ term to a Lennard-Jones interaction potential as follows:
\begin{eqnarray}
V(r) =  -4 \epsilon \Bigl ( \bigl({{\sigma} \over r}\bigr)^6
-\bigl({{\sigma} \over r}\bigr)^{12} \Bigr ) +
\alpha V_C(L,r)
\end{eqnarray}
where $V_C(L,r)$ is the $1/r$ modified to include the influence of the
periodic image charges in a finite-size $L \times L$ lattice  with periodic 
boundary conditions. This is implemented for two-dimensions 
by the well-known Ewald summation method\cite{ewald}.
For any given value of the parameter $\alpha$ we adjust the
value of $\epsilon$ in order for the minimum of the Lennard-Jones
potential to be at the same value $V_{min}=10^{\circ} K$. This is the 
approximate  minimum value of the Aziz potential\cite{aziz} 
that describes
the helium-helium interaction and it is used to study neutral helium
quantum films. The total interaction potential is shown in Fig.~\ref{fig0}
for various values of the parameter $\alpha$. Notice that
in order to keep $V_{min}$ a constant,
increasing the value of $\alpha$ requires an increased value of 
$\epsilon$ and this leads to a harder core potential and a more 
attractive force dragging the neighbor to toward the potential 
minimum. 
  
We will use the path-integral Monte Carlo method (PIMC) for superfluid
films as developed in Ref.\cite{pierce}. 
We have carried out simulations at $T=0.5^{\circ} K$ and $T=1^{\circ} K$ 
for all
100 combinations of 10 density values starting from
$\rho=0.0120\AA^{-2}$ up to $0.0480 \AA^{-2}$ with 
10 values of $\alpha$ starting from 
$\alpha=50^{\circ} K \AA$ up to 
$\alpha=500^{\circ} K \AA$ in increments of $50^{\circ} K \AA$.
We kept the number of particles fixed at 30 and we varied the size of
the simulation cell in order to vary the density.
We have computed the pair distribution function, and contour plots of the
particle probability density for all of the above parameter values.
Approximately 10,000  Monte Carlo sweeps were used for
thermalization and about 10,000 sweeps for 
measurements of observable.

\section{RESULTS}

\begin{figure}
\begin{center}
\includegraphics[width=\figwidth]{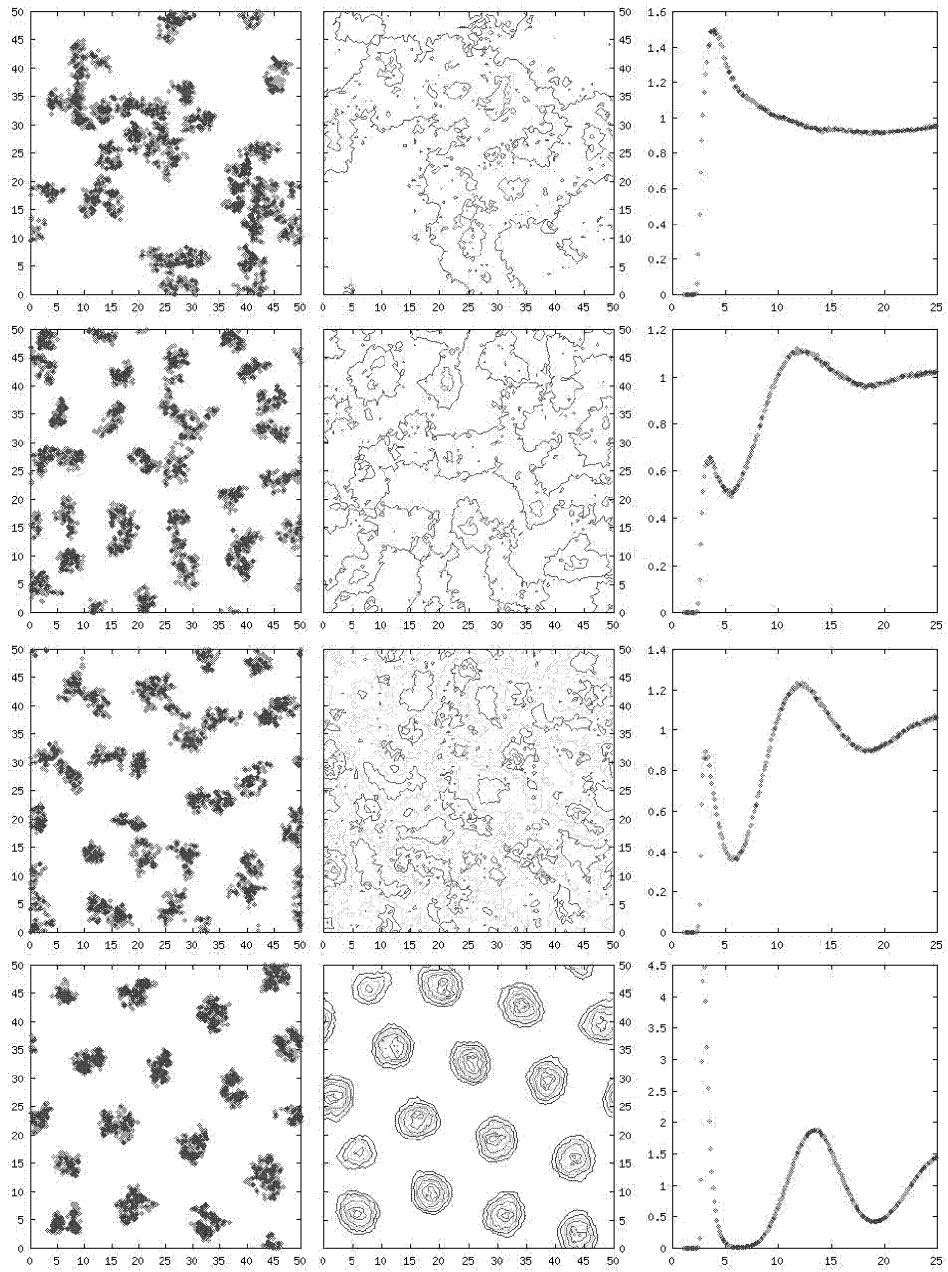}
\end{center}
\caption{A thermalized space-time configuration, the contour plot and the 
pair distribution at surface density 
$0.0120 \AA^{-2}$, $T= 0.5^{\circ} K$  and 
for $\alpha=0$ (top row) $\alpha=50^{\circ} K \AA$ 
(second row from the top) $\alpha=100^{\circ} K \AA$
(third from top row) and $\alpha=350^{\circ} K \AA$ (fourth row from the top)}
\label{fig1}
\end{figure}
In Fig.~\ref{fig1} we give a thermalized space-time configuration
of the system, contour plots of the probability distribution averaged over 
10,000 such configurations and  the pair distribution function for surface 
density $\rho=0.0120 \AA^{-2}$ and temperature $T=500 mK$ for 
various values of $\alpha$.
We notice that while for this surface density in the absence of the Coulomb 
repulsion the system phase separates, the presence of even weak Coulomb 
interaction tends to segregate the particles into clusters which spread 
over the  entire film surface. In addition,
the pair distribution function develops a second peak at the average 
inter-cluster distance. There is still a peak near the minimum of the
Van der Waals attractive part of the interaction but for small $\alpha$ 
this peak is much smaller
compared to that of the neutral fluid. Notice that as we raise the strength 
of the Coulomb repulsion the system forms a triangular superlattice of
clusters. The pair distribution function has a sharp peak near the minimum
of the Van der Waals attraction and an oscillatory behavior with a period
equal to the distance between nearest neighbor clusters. 

\begin{figure}
\begin{center}
\includegraphics[width=\figwidth]{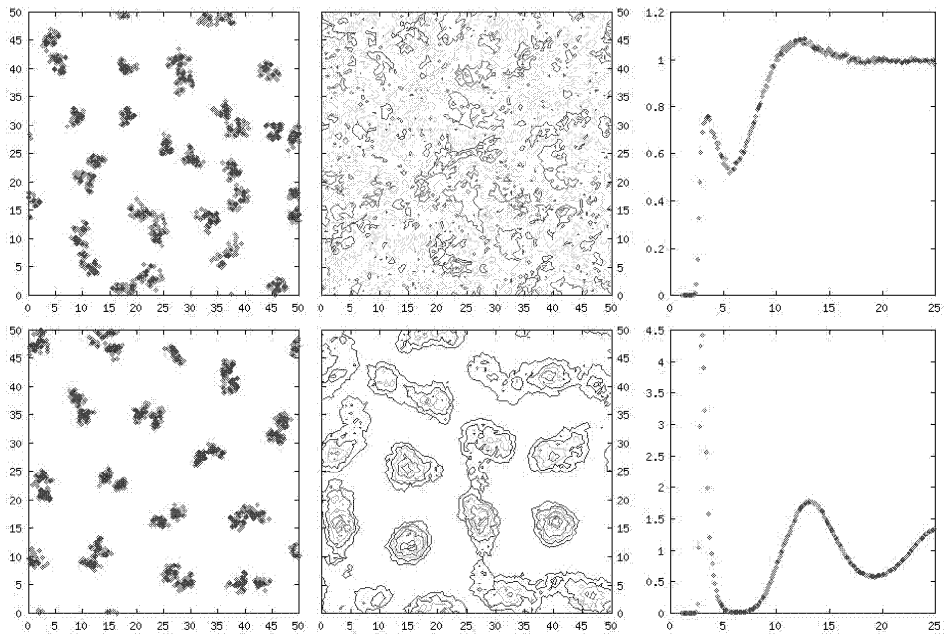}
\end{center}
\caption{A thermalized space-time configuration, the contour plot and the 
pair distribution at surface density  $0.0120 \AA^{-2}$, $T= 1^{\circ} K$ 
and $\alpha=50^{\circ} K \AA$ (top row) and $\alpha=350^{\circ} K \AA$ 
(bottom row)}
\label{fig2}
\end{figure}
\begin{figure}
\begin{center}
\includegraphics[width=\figwidth]{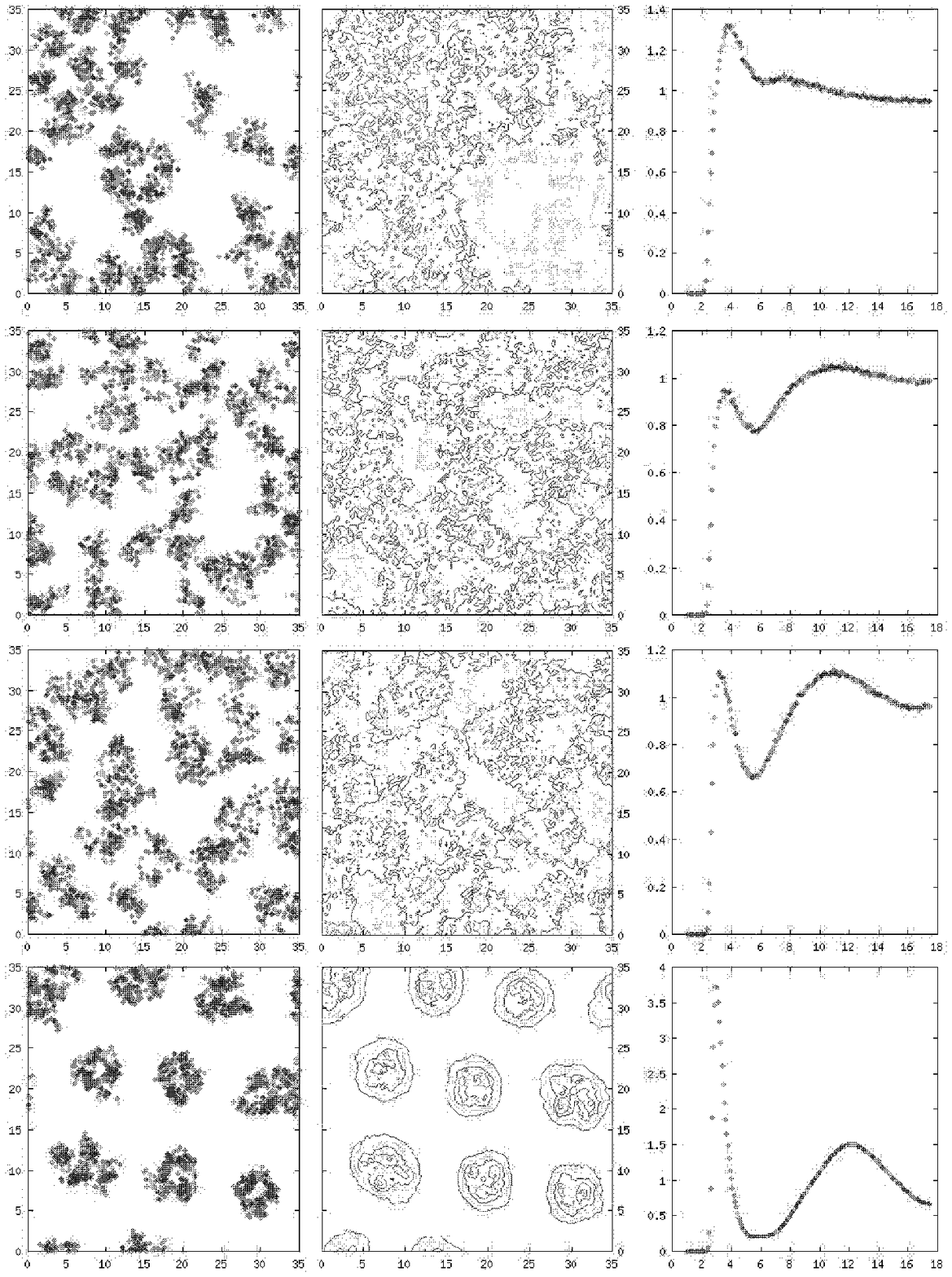}
\end{center}
\caption{A thermalized space-time configuration, the contour plot and the 
pair distribution at surface density  $0.0245 \AA^{-2}$, $T= 500 mK$ 
and $\alpha=0$ (top row), $\alpha=50 ^{\circ} K \AA$ (second from the top row),
$\alpha=100^{\circ} K \AA$ and $\alpha=300^{\circ} K \AA$ (bottom row)}
\label{fig3}
\end{figure}
\begin{figure}
\begin{center}
\includegraphics[width=\figwidth]{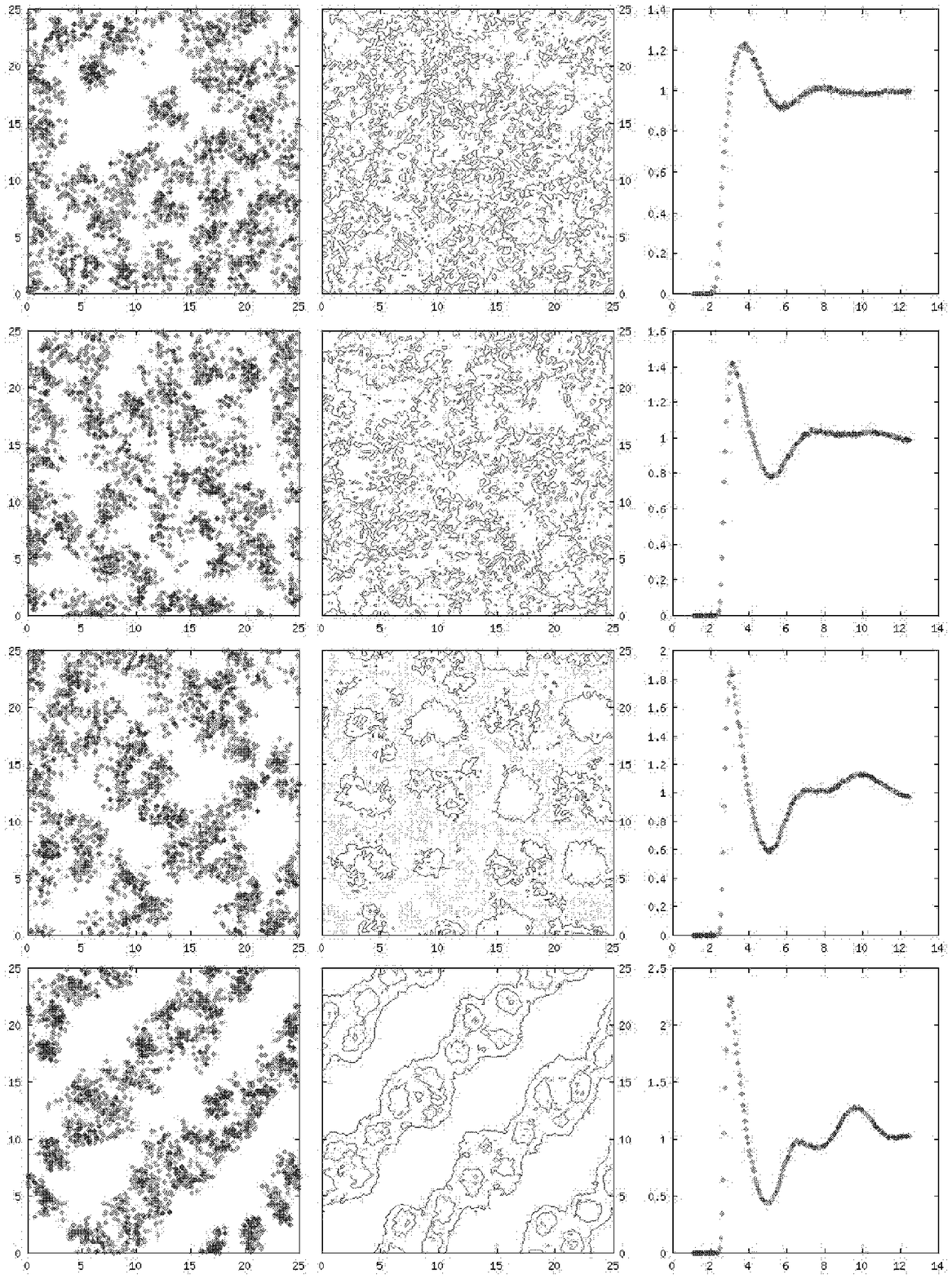}
\end{center}
\caption{A thermalized space-time configuration, the contour plot and the 
pair distribution at surface density  $0.0480 \AA^{-2}$, $T= 500 mK$ 
and $\alpha=0$ (top row), $\alpha=200^{\circ} K \AA$ (second from the top row),
$\alpha=300^{\circ} K \AA$ and $\alpha=350^{\circ} K \AA$ (bottom row)}
\label{fig5}
\end{figure}
\begin{figure}
\begin{center}
\includegraphics[width=\figwidth]{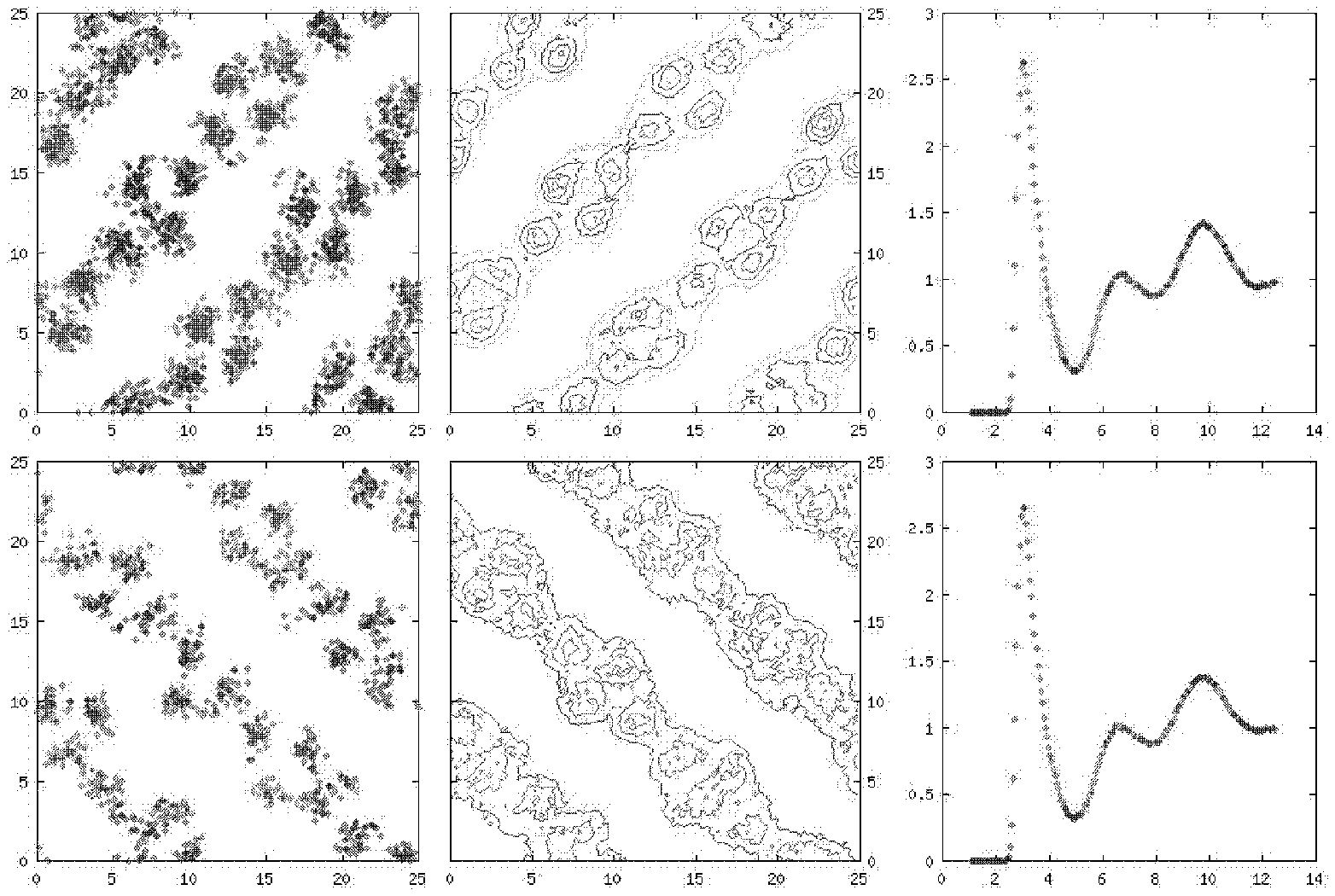}
\end{center}
\caption{A thermalized space-time configuration, the contour plot and the 
pair distribution at surface density  $0.0480 \AA^{-2}$, 
 $\alpha=450^{\circ} K \AA$ and $T=500 mK$ (top row), $T=1000 mK$ (bottom row)}
\label{fig6}
\end{figure}
By raising the temperature to $1^{\circ}K$
the picture remains similar as shown in Fig.~\ref{fig2}
with the tendency for the clusters and the particles to fluctuate more 
which will lead to the melting of the clusters at higher temperature. 
It is not 
clear if the system will become an ordered superlattice of such clusters
or will remain a fluid at zero temperature at small values of 
$\alpha$. It is well known that, 
strictly speaking, a two-dimensional system such as ours cannot break
a continuous symmetry and crystallize at any finite temperature; 
however, a finite-size system  may appear as ordered
because the correlation length for low enough temperature becomes
larger than the system size. 
 We have observed that, as a function 
of the simulation time, a finite-size 2D system drifts as a whole. As a result,
while within a relatively short time-scale the system appears ordered if we 
average configurations over a much larger time scale, the system will appear
disordered in a contour plot. This time scale depends on the value of 
$\alpha$. For small values of $\alpha$ it is easier for the system to
migrate as a whole while keeping the particle correlations. This is 
demonstrated in the figures for $\alpha=50,100^{\circ} K \AA$ where, while
the particle correlations show that they are those of a Wigner crystal
of clusters, the contour plots, taken over 10,000 configurations,
indicate that the system is uniform. The larger values of $\alpha$ require
averaging over much larger number of configurations to see the effects of the
drift discussed previously. 
\begin{figure}
\begin{center}
\includegraphics[width=3.5 in]{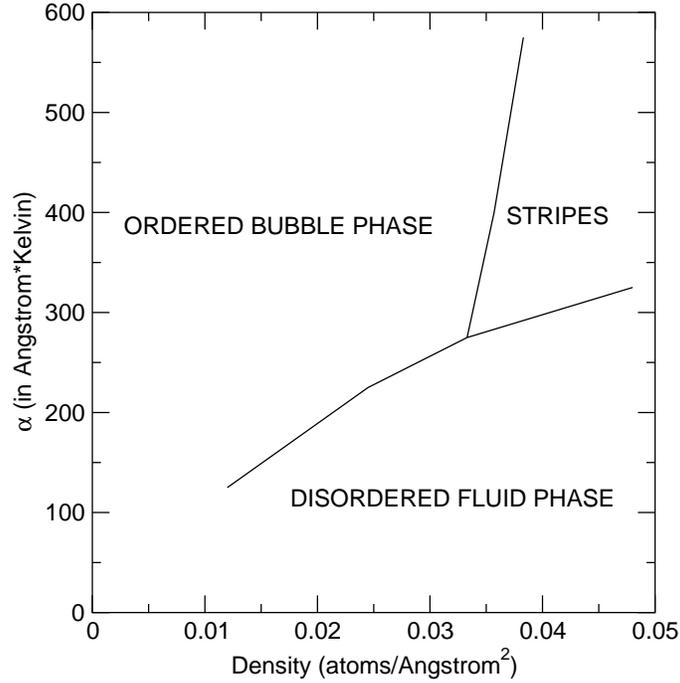}
\end{center}
\caption{The phase diagram of a submonolayer  
Van der Waals-Bose quantum film where an additional long-range Coulomb interaction
has been added.  }
\label{fig7}
\end{figure}

In Fig.~\ref{fig3} a thermalized snapshot of a space-time configuration
along with a contour plot of the probability distribution and the 
pair distribution is shown for a higher density of $0.0245 \AA^{-2}$, 
$T=500 mK$ and
various values of the Coulomb repulsion parameter $\alpha$. 
The situation is similar to that of density $\rho=0.0120 \AA^{-2}$.
We note, however,  that while the inter-cluster distance did not 
change significantly the
number of atoms in each cluster has increased by approximately the 
factor of the density increase.

In Fig.~\ref{fig5} we present the same three functions for 
$\rho=0.0480 \AA^{-2}$ 
which is very close to the equilibrium density of the neutral liquid.
Low values of the strength of Coulomb repulsion term have small influence
on these functions including the pair distribution function. We need to 
increase to a value $\alpha=200^{\circ} K \AA$ in order to see 
an influence of the 
repulsion on $g(r)$ comparable to that occurring at 
$\alpha=50^{\circ} K \AA$ for 
lower values of density. The reason for this insensitivity is the fact at
the equilibrium density the compressibility of the fluid is finite, while that
of the phase separated state is close to zero.

Notice as the values of $\alpha$ increases the clustering that occurs
is clusters of vacancies which tend to order on a triangular superlattice.
We have not clearly observed an ordered superlattice of clusters of 
vacancies. This tendency for ordering occurs near the equilibrium density
where in a phase separated state there is much more fluid than vacant space.
The Coulomb repulsion tends to compress and spread the fluid equally in all 
directions thus creating clusters of particles
or of vacancies or stripes. The stripes occur at densities near the 
equilibrium density and for large values of $\alpha$. Notice that $g(r)$
has three peaks when the stripes form or tend to form in Fig.~\ref{fig5}.
The short-distance peak is that which corresponds to the Van der Waals minimum,
the peak at the longest distance corresponds to the inter-stripe distance and
the intermediate peak corresponds to the a further ordering within the stripe.
Notice in the contour plots of Fig.~\ref{fig5} that the stripes are 
formed by merging of clusters and the stripe width is modulated along the
stripe and the modulation periodicity defines this third peak structure
of the pair distribution function.

In Fig.~\ref{fig6} we compare the situation for a larger value of 
$\alpha=450^{\circ} K \AA$ for the equilibrium density and at temperature $T=500 mK$ and
$T=1000 mk$. Notice that the detailed ordering of particles with a stripe
begins to go away at the higher temperature because of thermal fluctuations.
However, the peaks of the pair distribution function remain unchanged.

\section{PHASE DIAGRAM-CONCLUSIONS}
Our results are summarized on the phase diagram shown in Fig.~\ref{fig7}.
Within our large error bars in the determination of the phase boundaries
of the three phases, we have not been able to find any significant shift in the
phase boundaries with temperature.
The phase boundaries separate a uniform fluid phase from a phase of 
bubbles or clusters ordered in a triangular lattice formation
and a phase of ordered
stripes at high surface density and relatively strong Coulomb 
repulsion strength.

In order to draw conclusions relevant to the case of cuprate 
superconductors we need to also study the superfluid density
of the system as a function of the density and the Coulomb repulsion.
Due to the non-local nature of the permutation cycles and
the winding number operator this requires a significantly larger
computational time scales than the calculation presented here and
we are in the process of carrying it out.

\section*{Acknowledgements}

This work was supported by NASA grants NAG-1773 and NAG-2867.


\begin{thebibliography}{0}
\bibitem{emery} V. J. Emery, S. A. Kivelson, and H. Q. Lin, 
{\it Phys. Rev. Lett.}, {\bf 64}, 475 (1990). 
 S. A. Kivelson, E. Fradkin and V. J. Emery,
 Nature (London) {\bf 393}, 550 (1998).

\bibitem{hm} C. S. Hellberg and E. Manousakis, Phys. Rev. {\bf B 61}, 
11787 (2000).  
 C. S. Hellberg and E. Manousakis, Phys. Rev. Lett. {\bf 78}, 
4609 (1997). 

\bibitem{carlson}E. W. Carlson, V. J. Emery, S. A. Kiveslon, D. Ograd, 
``The Physics of Conventional and Unconventional Superconductors'' ed. 
by K. H. Bennemann and J. B. Ketterson (Springer-Verlag).

\bibitem{bm}M. Boninsegni and E. Manousakis
Phys. Rev. B 47, 11897-11904 (1993) 
\bibitem{manousakis03} 
E. Manousakis, Phys. Rev.  B 67, 195103 (2003).

\bibitem{greywall93}
D.~S. Greywall, Phys. Rev. B {\bf 47},  309  (1993).

\bibitem{reppy93}
P.~A. Crowell and J.~D. Reppy, Phys. Rev. Lett. {\bf 70},  3291  (1993).

\bibitem{reppy96}
P.~A. Crowell and J.~D. Reppy, Phys. Rev. B {\bf 53},  2701  (1996).

\bibitem{pierce} M.~Pierce and E.~Manousakis, Phys. Rev. Lett. {\bf 81}, 
 156 (1998). M. ~Pierce and E. ~Manousakis, Phys. Rev. B {\bf 59},
3802 (1999). M. ~Pierce and E. ~Manousakis, Phys. Rev. B {\bf 62},
5228 (2000). M. ~Pierce and E. ~Manousakis, Phys. Rev. Lett. {\bf 83},
5314 (1999).

\bibitem{carlos}
W. ~E. Carlos and M. ~W. Cole, Surf. Sci. {\bf 91}, 339 (1980).
\bibitem{aziz}
R.~A. Aziz {\it et~al.}, Mol. Phys. {\bf 77},  321  (1992).


\bibitem{ewald} D. M. Ceperley, Phys. Rev. {\bf B 18}, 3126 (1978).


\end{thebibliography}
\end{document}